\documentclass[sigconf]{acmart}

\usepackage{booktabs}
\usepackage{xurl}
\newcommand{\code}[1]{\path{#1}}
\usepackage{makecell}

\usepackage{pgfplots}
\pgfplotsset{compat=1.18}
\usepackage{pgfplotstable}
\usepgfplotslibrary{colormaps}

\usepackage{tikz}
\usetikzlibrary{arrows.meta, positioning, shapes.geometric}

\AtBeginDocument{%
  }

\copyrightyear{2026}
\acmYear{2026}
\setcopyright{cc}
\setcctype{by}
\acmConference[CSCW Companion '26]{Companion of the Computer-Supported Cooperative Work and Social Computing}{October 10--14, 2026}{Salt Lake City, UT, USA}
\acmBooktitle{Companion of the Computer-Supported Cooperative Work and Social Computing (CSCW Companion '26), October 10--14, 2026, Salt Lake City, UT, USA}
\acmDOI{10.1145/3785651.3831560}
\acmISBN{979-8-4007-2378-0/2026/10}

\begin{document}

\title{Who Acts When the User Is Gone? Digital Remains, Survivor Claims, and Post-Mortem Governance}

\author{Supriya Khadka}
\affiliation{%
  \institution{George Mason University}
  \city{Fairfax, Virginia}
  \country{USA}}
\email{skhadk@gmu.edu}

\author{Dhiman Goswami}
\affiliation{%
  \institution{George Mason University}
  \city{Fairfax, Virginia}
  \country{USA}}
\email{dgoswam@gmu.edu}

\author{Sanchari Das}
\affiliation{%
  \institution{George Mason University}
  \city{Fairfax, Virginia}
  \country{USA}}
\email{sdas35@gmu.edu}

\renewcommand{\shortauthors}{Khadka et al.}

\begin{abstract}
Digital systems continue to govern accounts, devices, data, and recovery channels after an account holder dies, leaving survivors to manage digital remains through mechanisms built around a living user. We examine post-mortem digital governance as a sociotechnical problem of cooperative and contested work, focusing on who acts when the user is gone, what claims they make, and what barriers shape recovery, preservation, closure, and protection. We conducted a content analysis of $800$ Reddit posts about post-mortem digital privacy and security, coding posts across assets, actors, actions, privacy tensions, access barriers, policy gaps, emotional contexts, and risks. Findings show that phones/devices often act as gateways to other digital remains, socially connected actors make most claims, and data loss emerges as a central harm. We synthesize these findings into a Post-Mortem Digital Governance Framework for designing mechanisms that support survivor coordination while limiting access by purpose, asset, actor, and context.
\end{abstract}




\maketitle

\section{Introduction}
When an account holder dies, the digital systems connected to them often remain active. Accounts, devices, data, and recovery channels may still require decisions about access, preservation, closure, and protection while continuing to operate under assumptions built for an active user. A locked phone may contain family photos, messages, or account recovery codes that survivors cannot otherwise retrieve. A social media profile may continue appearing in notifications, become a space for remembrance, or face unwanted access. Email, cloud storage, financial accounts, subscriptions, and password managers remain governed by platform policies, billing cycles, fraud controls, and recovery procedures~\cite{acker2014death,brubaker2016legacy,doyle2023digital,locasto2011security, markert2023transcontinental, walsh2021my}.

For surviving family members, partners, friends, executors, and other affected actors, death can transform ordinary security mechanisms into urgent problems of privacy, memory, fraud, evidence, and estate management~\cite{morse2022posthumous,holt2021personal,harbinja2017post}. These situations expose a boundary condition for account-centric security, since the systems persist, but the user can no longer participate in decisions about them~\cite{acquisti2015privacy,doyle2023digital}. The resulting work is cooperative and contested, requiring survivors to navigate family relationships, platform rules, legal procedures, institutional processes, and technical recovery mechanisms while making claims over digital remains.

We use \textit{post-user security} to describe security and privacy problems that emerge when digital systems continue operating after the account holder can no longer act. Prior work on digital legacy, posthumous privacy, and account recovery has established the normative and legal challenges surrounding post-mortem data, often examining issues through institutional categories such as heirs, fiduciaries, consent, dignity, property, and platform terms~\cite{harbinja2017post,buitelaar2017post,tkach2024model,patti2019digital}. Our work complements this literature by focusing on the operational breakdowns that arise when these assumptions continue to govern accounts after the user is no longer able to participate. We use the term \textit{post-user} to emphasize that the central problem is not death itself, but the absence of the user while the system continues to assume their availability. Systems continue to expect authentication, consent, objections, and clarification of intent, even though the account holder can no longer provide them. We examine these challenges through the lens of post-mortem digital governance, which encompasses the social, technical, and institutional arrangements that determine who can act, what actions are possible, and whose interests are protected. In practice, however, acting on behalf of a deceased user remains difficult because social responsibility, legal authority, technical access, and platform policy often fail to align~\cite{buitelaar2017post,park2020ontology,slaughter2015barriers,ulguim2018digital}.

To understand how post-mortem digital governance becomes cooperative and contested work among survivors, platforms, and institutions, we ask:

\begin{itemize}
    \item What digital assets become privacy and security problems after death?
    \item Who attempts to act, and what kinds of post-mortem claims do they make?
    \item What barriers, risks, and policy gaps shape post-mortem digital governance?
\end{itemize}

We contribute an empirical analysis of 800 Reddit posts and introduce the \textbf{Post-Mortem Digital Governance Framework}. The framework reframes post-mortem digital privacy and security as a governance problem involving digital remains, claiming actors, post-mortem actions, governance tensions, access barriers, policy gaps, emotional context, and risk.

\section{Background: Digital Remains and Post-Mortem Governance}
Prior CSCW and HCI work shows that digital data remains socially, emotionally, legally, and economically significant after death. Researchers have examined memorial profiles~\cite{acker2014death,jiang2018tending,brubaker2016legacy,brubaker2011we,brubaker2019orienting}, personal archives~\cite{orita2022inheritance,akramov2024impact}, inheritance and estate management~\cite{tkach2024model,patti2019digital,farooqui2022inheritance}, and medical or genetic records~\cite{maixner2001confidentiality,siminoff2017confidentiality,shade2019ethical}. This literature identifies tensions between heir access and deceased privacy, legal authority and platform terms, and individual autonomy and family benefit~\cite{pierer2019inheritability,harbinja2023governing,park2020ontology,viana2017analysis, hutt2023right, adhikari2023evolution}.

We build on this work by focusing on the everyday operational breakdowns that occur when systems continue to operate after the user can no longer authenticate, reset credentials, approve recovery, appeal decisions, or clarify intent. These problems often center on specific actions such as recovering photos, closing accounts, exporting files, preserving evidence, preventing fraud, memorializing profiles, or planning future disclosure~\cite{kutler2011protecting,andreea2026post,allen2024postmortem,thangaraj2025s}. We therefore treat post-mortem digital governance as a sociotechnical problem shaped by technical mechanisms, social relationships, emotional stakes, and institutional processes~\cite{doyle2023digital,holt2021personal,morse2022posthumous,buitelaar2017post}.

Post-mortem digital governance often involves distributed work among relatives, friends, partners, executors, platforms, legal institutions, and financial services. These actors must make decisions under grief, uncertainty, incomplete authority, missing credentials, and conflicting privacy expectations~\cite{brubaker2014stewarding}. In this paper, we treat digital remains as objects around which social and institutional work is organized.

\section{Method}
We conducted a qualitative content analysis of 800 public Reddit posts about post-mortem digital privacy and security. Reddit provides rich discussions of these issues across everyday advice-seeking communities, including technical support, legal advice, estate planning, personal finance, grief support, relationships, privacy, and general advice forums.

We constructed the corpus through a multi-stage search and screening process. We began with exploratory searches to identify communities where post-mortem digital issues recur. Because these discussions are not confined to privacy-focused forums, we selected ten subreddits spanning technical support, legal advice, estate planning, personal finance, grief support, relationships, privacy, and general advice. We prioritized topical relevance over community size, treating the corpus as a source of recurring sociotechnical patterns rather than a representative sample; consequently, smaller communities contribute fewer posts. We then searched these subreddits using mortality-related and digital-account terms, including \texttt{digital legacy}, \texttt{dead man switch}, \texttt{legacy contact}, \texttt{passed away password}, \texttt{deceased account}, and \texttt{dead person phone}, yielding 3,990 candidate posts.

We manually screened posts against two criteria, requiring both a mortality or post-mortem context and a digital, privacy, security, or data-governance issue. The final corpus contained 800 relevant posts from 2009--2026. Each case was coded across eight dimensions: \code{digital_asset}, \code{actor_requesting_access}, \code{desired_action}, \code{privacy_tension}, \code{barrier}, \code{policy_gap}, \code{emotional_context}, and \code{risk_type}. Coding combined deductive categories derived from our research questions with inductive refinement during screening and initial coding, following qualitative content analysis and constant-comparison approaches~\cite{hsieh2005three,glaser1965constant, naveen2026privacy}. When multiple labels were plausible, we assigned the dominant label representing the central issue. A single researcher coded the corpus, and the codebook was reviewed by a second researcher before coding began. Ambiguous cases were revisited after the initial coding pass.

We computed descriptive frequencies for each coding dimension and examined recurring patterns across them. We use these patterns descriptively to characterize how post-mortem digital governance appears in the corpus rather than to estimate population-level prevalence. Because the corpus concerns death, grief, family conflict, fraud, and sensitive personal data, we treated the posts as sensitive user-generated content. We did not retain usernames in the reporting dataset, report findings only in aggregate, and do not publish raw post text, URLs, or post identifiers. When examples informed our interpretation, we paraphrased them to reduce searchability and re-identification risk.

\section{Findings}
We organize the findings around our three research questions. Section 4.1 addresses which digital assets become problems (RQ1), Section 4.2 addresses who acts and what claims they make (RQ2), and Sections 4.3 and 4.4 address the harms and governance gaps that shape post-mortem governance (RQ3). Table~\ref{tab:top-frequencies} summarizes the most frequent labels.

\begin{table}[t]
\centering
\small
\setlength{\tabcolsep}{4pt}
\renewcommand{\arraystretch}{0.95}
\caption{Most frequent labels across key coding dimensions.}
\label{tab:top-frequencies}
\begin{tabular}{@{}>{\raggedright\arraybackslash}p{0.24\linewidth}p{0.70\linewidth}@{}}
\toprule
\textbf{Dimension} & \textbf{Most frequent labels} \\
\midrule
Digital asset & Phone/device: 239 (29.9\%); Social media profile: 135 (16.9\%); Financial account: 100 (12.5\%). \\
\addlinespace
Actor & Child: 229 (28.6\%); Extended family: 122 (15.2\%); Self-planning: 109 (13.6\%). \\
\addlinespace
Desired action & Recover: 169 (21.1\%); Access: 141 (17.6\%); Plan ahead: 98 (12.2\%). \\
\addlinespace
Barrier & Locked device/account: 157 (19.6\%); Platform policy: 98 (12.3\%); Technical recovery issue: 96 (12.0\%). \\
\addlinespace
Policy gap & Device access unsupported: 243 (30.4\%); No digital legacy plan: 109 (13.6\%); Platform process unclear: 105 (13.1\%). \\
\addlinespace
Risk type & Data loss: 393 (49.1\%); Financial loss: 126 (15.8\%); None/unclear: 106 (13.2\%). \\
\bottomrule
\end{tabular}
\end{table}

\subsection{Digital remains span devices, accounts, and records}
The most common digital asset was the deceased person's phone or device (Table~\ref{tab:top-frequencies}), followed by social media profiles, financial accounts, email or cloud accounts, photos or videos, credentials, and messages. Post-mortem governance therefore spans local devices, platform accounts, financial systems, communication records, and personal media.
Phones and devices often acted as gateways to other digital remains~\cite{fernandes2023you}. Authentication apps, recovery prompts, and saved credentials resided on the device, so phones sat upstream of the accounts they unlocked. A single locked handset could trigger a chain of downstream account failures that no individual platform policy could resolve, and device-level barriers shaped what survivors could attempt before any platform process became available. In device-related posts, survivors most often sought access and recovery, and data loss was the dominant risk. Governance frequently began with a physical object in the home. Phones, laptops, hard drives, password notebooks, and recovery devices all became focal points for family coordination and access to institutions.

\subsection{Post-mortem claims come from socially connected actors}
The actor attempting to act after death was usually someone with a social or practical connection to the deceased. Children were the largest actor category, followed by extended family, self-planning cases, unknown actors, friends, siblings, spouses or partners, and parents. Executors or legal representatives appeared in only a small share of posts. These patterns show that post-mortem governance is shaped by relational responsibility as well as formal authority.

The claims made by these actors varied. Children often described managing a parent's devices, accounts, bills, or memories. Siblings and extended family described recovering data, preserving photos, or resolving account issues. Friends and partners appeared in cases involving online traces, messages, memorialization, evidence, or emotional connection. Self-planning cases involved users deciding what should happen to their accounts, passwords, files, or messages after death. 

Beyond who appears, these actors made different kinds of claims. Practical claims involved managing bills, closing accounts, recovering files, or completing estate tasks; emotional claims involved preserving photos, messages, or voice recordings; protective claims focused on preventing fraud or securing accounts; and evidentiary claims sought to preserve or recover messages after suspicious or contested circumstances. These claims often collided within a single case, such as when a child sought to close an account while a partner wanted to preserve its messages. The design question therefore concerns not only which actors are admitted, but also which claims a given action serves. These findings show why systems cannot rely solely on formal roles such as executors or pre-designated legacy contacts.

\subsection{Data loss is a central post-mortem harm}

We use \textit{post-mortem harm} to mean a setback to the interests of the deceased, survivors, or third parties that is produced or left unaddressed by how a system behaves once the user can no longer act. Conventional privacy and security research has largely focused on harms associated with confidentiality failures, including exposure, unauthorized access, disclosure, and misuse~\cite{acquisti2015privacy}. Our corpus highlights an additional class of harms centered on availability and stewardship.

Data loss emerged as the dominant risk in the corpus. Posts described survivors unable to retrieve photos or messages from locked devices, recover accounts because a recovery phone or authentication app was unavailable, manage accounts after death, or preserve records with family, financial, or evidentiary value. For survivors, data loss was concrete and emotionally significant.

These findings broaden how post-mortem harms are understood. In post-mortem settings, harm can arise from permanent loss, blocked preservation, unresolved obligations, or failed evidence protection alongside the confidentiality concerns that have traditionally received greater attention. Survivor access should still be carefully limited, since disclosure can create privacy and abuse risks. At the same time, systems that prevent unauthorized access while also preventing legitimate preservation can create additional harms. Preservation, transfer, closure, and selective access therefore become part of the cooperative work of post-mortem governance.

\begin{figure*}[t]
\centering
\resizebox{\textwidth}{!}{%
\begin{tikzpicture}[
    node distance=0.45cm and 0.55cm,
    every node/.style={font=\scriptsize, align=center},
    start/.style={
        rectangle, rounded corners, draw=blue!60!black, thick, fill=blue!10,
        text width=2.45cm, minimum height=1.05cm, inner sep=4pt
    },
    dim/.style={
        rectangle, rounded corners, draw=teal!60!black, thick, fill=teal!10,
        text width=1.95cm, minimum height=0.9cm, inner sep=3pt
    },
    context/.style={
        rectangle, rounded corners, draw=orange!70!black, dashed, thick, fill=orange!12,
        text width=3.0cm, minimum height=0.8cm, inner sep=3pt
    },
    outcome/.style={
        rectangle, rounded corners, draw=green!55!black, thick, fill=green!12,
        text width=2.65cm, minimum height=1.05cm, inner sep=4pt
    },
    arrow/.style={-{Latex[length=1.6mm]}, thick, draw=blue!70!black},
    softarrow/.style={-{Latex[length=1.6mm]}, thick, draw=gray!65}
]

% Main horizontal sequence
\node[start] (postuser) {
\textbf{Post-user condition}\\
User can no longer authenticate, consent, recover access, object, or clarify intent
};

\node[dim, right=of postuser] (asset) {
\textbf{Digital\\remains}\\
What persists?
};

\node[dim, right=of asset] (actor) {
\textbf{Claiming\\actors}\\
Who acts?
};

\node[dim, right=of actor] (action) {
\textbf{Post-mortem\\actions}\\
What is sought?
};

\node[dim, right=of action] (tension) {
\textbf{Governance\\tensions}\\
What values conflict?
};

\node[dim, right=of tension] (barrier) {
\textbf{Access\\barriers}\\
What blocks resolution?
};

\node[dim, right=of barrier] (gap) {
\textbf{Policy\\gaps}\\
What is unsupported?
};

\node[outcome, right=of gap] (governance) {
\textbf{Post-mortem\\digital governance}\\
Purpose-limited, asset-specific, actor-sensitive, context-aware
};

% Cross-cutting layers
\node[context, above=0.75cm of action] (emotion) {
\textbf{Emotional context}\\
grief, memorialization, estate administration, family conflict, suspicious death
};

\node[context, below=0.75cm of action] (risk) {
\textbf{Risk type}\\
data loss, fraud, privacy violation, financial loss, legal risk, emotional harm
};

% Main arrows
\draw[arrow] (postuser) -- (asset);
\draw[softarrow] (asset) -- (actor);
\draw[softarrow] (actor) -- (action);
\draw[softarrow] (action) -- (tension);
\draw[softarrow] (tension) -- (barrier);
\draw[softarrow] (barrier) -- (gap);
\draw[arrow] (gap) -- (governance);

% Cross-cutting influence arrows
\draw[arrow] (emotion.south) -- (action.north);
\draw[arrow] (emotion.south west) to[bend right=10] (actor.north);
\draw[arrow] (emotion.south east) to[bend left=10] (tension.north);

\draw[arrow] (risk.north) -- (action.south);
\draw[arrow] (risk.north east) to[bend right=10] (tension.south);
\draw[arrow] (risk.east) to[bend right=12] (gap.south);

\end{tikzpicture}%
}
\caption{Post-Mortem Digital Governance Framework. }
\Description{A left-to-right flow diagram with nine boxes. At the far left, a blue box labeled ``Post-user condition'' states that the user can no longer authenticate, consent, recover access, object, or clarify intent. An arrow leads right into a chain of six teal boxes, each posing a question: Digital remains (what persists?), Claiming actors (who acts?), Post-mortem actions (what is sought?), Governance tensions (what values conflict?), Access barriers (what blocks resolution?), and Policy gaps (what is unsupported?). A final arrow leads to a green box labeled ``Post-mortem digital governance,'' described as purpose-limited, asset-specific, actor-sensitive, and context-aware. Two orange dashed boxes sit above and below the middle of the chain. The upper box, ``Emotional context,'' lists grief, memorialization, estate administration, family conflict, and suspicious death, with arrows pointing down into Claiming actors, Post-mortem actions, and Governance tensions. The lower box, ``Risk type,'' lists data loss, fraud, privacy violation, financial loss, legal risk, and emotional harm, with arrows pointing up into Post-mortem actions, Governance tensions, and Policy gaps.}
\label{fig:framework}
\end{figure*}
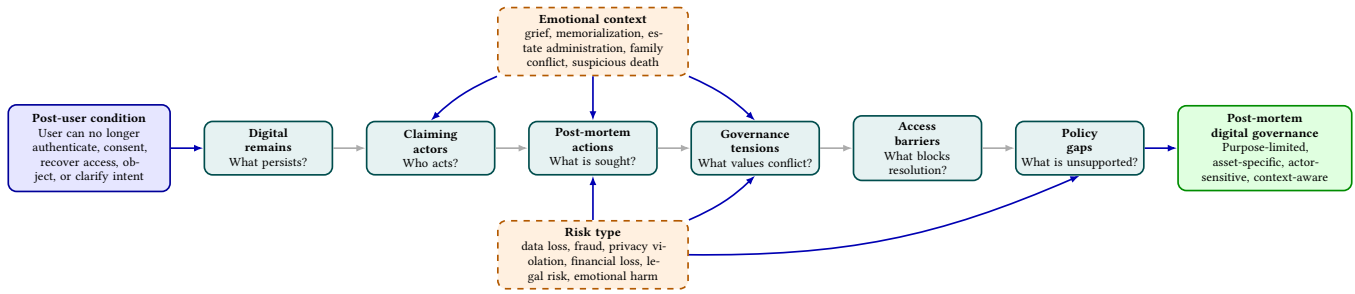

\subsection{Governance gaps are fragmented across systems}

The most common policy gap was unsupported device access, followed by absent digital legacy planning, unclear platform processes, fraud response gaps, platform unresponsiveness, and legal uncertainty. These gaps appeared across several layers, including devices and local data, platform account processes, planning tools, fraud response, and legal authority. This fragmentation means that survivors often face multiple systems with different rules, evidence requirements, timelines, and assumptions about who is allowed to act.

Post-mortem tasks often depended on sequences of access across people, devices, platforms, and institutions. A survivor may need the deceased person's phone to reach an email account, the email account to reset another service, a death certificate to close a financial account, or a platform-specific form to memorialize or remove a profile. These dependencies make post-mortem governance brittle, since failure at one point can block the following task. They also show why post-mortem governance cannot be solved through a single account setting or platform policy.

\section{Post-Mortem Digital Governance Framework}
We synthesize these findings into a Post-Mortem Digital Governance Framework (Figure~\ref{fig:framework}), which treats post-mortem digital governance as cooperative work around digital remains. Emotional context and risk type cut across its dimensions and help explain why a request matters and what harm is at stake. For CSCW researchers and designers, the framework shows where coordination can break down between assets and actions, social relationships and formal authority, emotional stakes and institutional procedures, and access barriers and policy support.

The framework supports four design commitments. First, governance should be \textit{purpose-limited}. Systems should distinguish read-only review, data export, account closure, fraud freeze, billing management, evidence preservation, memorialization, and trusted-contact notification rather than treating post-mortem requests as a single category. Each action names a beneficiary, which lets platforms scope disclosure to the interest a request actually serves. Second, governance should be \textit{asset-specific}. A phone, cloud archive, social media profile, financial account, and private message thread raise different risks and support needs. Third, governance should be \textit{actor-sensitive}. Legal representatives, designated contacts, children, partners, friends, and suspicious actors require different levels of scrutiny and support. Fourth, governance should be \textit{context-aware}~\cite{khadka2026riskcost}. The same request can carry different stakes in cases involving grief, estate administration, fraud, family conflict, or suspicious death.

\section{Discussion and Implications}
Our findings position post-user security as a stress test for account-centric security. This creates a governance gap between persistent digital systems and the absence of the person those systems were designed to serve.

For CSCW and sociotechnical systems research, this gap shows how digital remains become sites of cooperative and contested work. It extends prior accounts of memorialization and stewardship, which show how survivors care for and curate the profiles of the dead~\cite{brubaker2016legacy,brubaker2014stewarding,jiang2018tending}, as well as work on digital inheritance that examines the transfer of assets to heirs~\cite{tkach2024model,patti2019digital,farooqui2022inheritance}. That work assumes a survivor who has already reached the asset. Our corpus shows survivors confronting operational breakdowns in access and recovery before care, stewardship~\cite{brubaker2014stewarding}, or transfer can begin, and it shows those breakdowns distributed across relatives, platforms, and institutions holding uneven authority and knowledge~\cite{doyle2023digital,holt2021personal,buitelaar2017post}. This shifts the design focus from account recovery alone to the coordination work required when responsibility, authority, and access are distributed across multiple people and systems. Families, friends, partners, executors, platforms, and institutions may all become involved, though they do not share the same authority, knowledge, emotional relationship, or exposure to risk. Survivors pursue these actions while platforms and institutions must protect the deceased person, third parties, and the integrity of their systems. Post-mortem digital governance therefore requires support for coordination across actors whose responsibilities and permissions do not always align.

Planning tools such as legacy contacts, digital wills, inactive-account managers, and password managers can reduce uncertainty, though they cannot carry the whole system. Many cases involve absent planning, unclear instructions, inaccessible recovery channels, sudden death, or actors who are socially responsible without being formally designated. Post-mortem governance therefore needs cautious fallback pathways, including document-based review, limited-purpose requests, preservation holds, fraud freezes, delayed disclosure, logging, review, and appeal. These mechanisms should support legitimate post-mortem needs while reducing risks of surveillance, coercion, family conflict, and opportunistic account takeover~\cite{gupta2024really}.

\section{Conclusion}
When an account holder dies, the digital systems connected to them often remain active, but the person those systems are designed to rely on can no longer participate. Based on an analysis of 800 Reddit posts, we show that post-mortem digital privacy and security problems span devices, accounts, records, relationships, and institutions. We introduce post-user security to describe the security and privacy problems that emerge when digital systems outlast the user's ability to act, and we offer a Post-Mortem Digital Governance Framework for designing purpose-limited, asset-specific, actor-sensitive, and context-aware responses. These findings suggest that post-mortem digital governance should support coordinated work across survivors, platforms, and institutions while protecting the deceased, survivors, and third parties.

\section*{Acknowledgements}
We acknowledge the Data Agency and Security (DAS) Lab at George Mason University (GMU), and Google for partially supporting this work. The opinions expressed are solely those of the authors.

\section*{Generative AI Disclosure}
The authors did not use generative AI or large language model tools in the preparation of this submission.

\bibliographystyle{ACM-Reference-Format}
\bibliography{main}

\end{document}